\documentclass[conference]{IEEEtran}
\IEEEoverridecommandlockouts
\usepackage{cite}
\usepackage{amsmath,amssymb,amsfonts}
\usepackage{algorithmic}
\usepackage{graphicx}
\usepackage[symbol]{footmisc}

\usepackage{textcomp}
\usepackage{xcolor}
\def\BibTeX{{\rm B\kern-.05em{\sc i\kern-.025em b}\kern-.08em
    T\kern-.1667em\lower.7ex\hbox{E}\kern-.125emX}}
\graphicspath{ {figures/} }

\begin{document}

\title{Susceptibility of Adversarial Attack on Medical Image Segmentation Models}

\author{\IEEEauthorblockN{Zhongxuan Wang* \thanks{* Both authors contributed equally to this paper.}}
\IEEEauthorblockA{
\textit{Saint Andrew's School}\\
Boca Raton, United States of America \\
cndanielwang@gmail.com}
\and
\IEEEauthorblockN{Leo Xu*}
\IEEEauthorblockA{\textit{Lynbrook High School}\\
San Jose, United States of America\\
leoxu27@gmail.com }
}

\maketitle

\begin{abstract}
The nature of deep neural networks has given rise to a variety of attacks, but little work has been done to address the effect of adversarial attacks on segmentation models trained on MRI datasets. In light of the grave consequences that such attacks could cause, we explore four models from the U-Net family and examine their responses to the Fast Gradient Sign Method (FGSM)\cite{FSGM} attack.

We conduct FGSM attacks on each of them and experiment with various schemes to conduct the attacks. In this paper, we find that medical imaging segmentation models are indeed vulnerable to adversarial attacks and that there is a negligible correlation between parameter size and adversarial attack success. Furthermore, we show that using a different loss function than the one used for training yields higher adversarial attack success, contrary to what the FGSM authors suggested. In future efforts, we will conduct the experiments detailed in this paper with more segmentation models and different attacks. We will also attempt to find ways to counteract the attacks by using model ensembles or special data augmentations. Our code is available at https://github.com/ZhongxuanWang/adv\_attk
\end{abstract}


\section{Introduction}
\begin{figure}
    \centering
    \includegraphics[scale=0.35]{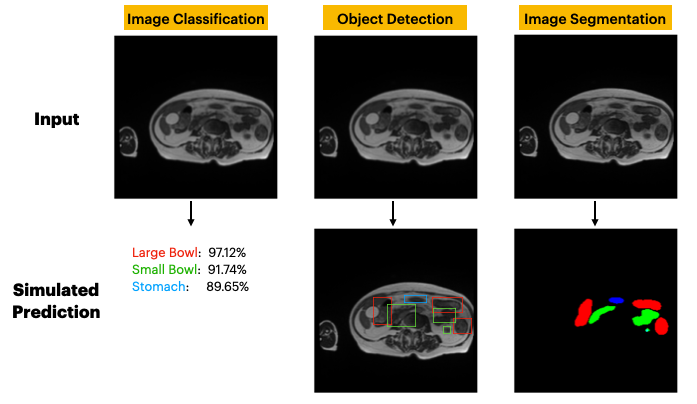}
    \caption{Difference between image classification, object detection, and image segmentation methods in the medical imaging field. Predictions are simulated except for the image segmentation method.}
    \label{fig:CVMethodsDiff}
\end{figure}

Today, deep convolutional neural networks (CNNs)\cite{CNN} have become increasingly popular in medical imaging, playing a role in the classification brain tumors, detection of organ boundaries, or segmentation of organ tumors. Since CNNs can exploit the spatial information present in images\cite{CNN}, they have been widely used in hospitals to provide doctors with valuable insights at an increased speed. Among all the medical imaging tasks, image segmentation is arguably one of the more challenging ones since it needs to leverage both global and local features to create masks for objects. While image classification helps doctors know the class of the image as a whole, and object detection helps doctors know the general location of the object, image segmentation allows doctors to see the boundaries of objects of interest clearly \cite{fritz.ai}. Their differences are illustrated in Figure \ref{fig:CVMethodsDiff}.

To leverage CNNs in biomedical image segmentation, Ronneberger et al. proposed U-Net, a revolutionary architecture that consists of a contraction path and a symmetric expansion path \cite{UNet}. Recently, many variants of U-Net have been proposed to achieve state-of-the-art (SOTA) performances in various medical imaging tasks. Examples of a few include nested U-Nets, U-Nets with dense skip connections to learn full-scale semantic information \cite{UNET2P} \cite{UNET3P}, and U-Net with transformer-based encoder or decoder to learn long-range semantic information \cite{TRANSUNET} \cite{SWINUNET}. These U-Net variants have all shown superior performances over the original U-Net.

Although the U-Net family possesses great potential in image segmentation for MRI data, recent studies have prioritized model performance over security. In fact, a variety of attacks have recently arisen to intentionally fool models into making incorrect predictions with high confidence by modifying the training dataset, testing dataset, model parameters, along others \cite{BACKDOORLEARNINGSURVEY}. Given the confidentiality of medical imaging datasets, it is usually impractical to poison the training dataset an MRI model was trained on or modify its parameters. Thus, poisoning inference data is a much larger concern for doctors. One of the methods for attacking inferencing data is known as a white-box adversarial attack, which assumes that attackers cannot modify the training data or the model but know about the model such as its architecture and weights.

In the context of medical image segmentation, successful adversarial attacks could incur hefty and irreversible consequences. For instance, if poisoned tumor segmentations misled doctors, doctors may overlook portions of tumor tissues that could cause death. If doctors relied on compromised tumor segmentation images to kill diseased tissues, both benign and vital parts of the organ may be damaged permanently. Unfortunately, most doctors are not trained to discern poisoned data from those unaffected, nor are they trained to take deterrent measures. To make the matter worse, Ma et al. pointed out that medical imaging models are more vulnerable to adversarial attacks than other types of models \cite{AdvMedical}. Thus, studying the effect of adversarial attacks on medical images has overarching significance.

\begin{figure}
    \centering
    \includegraphics[scale=0.27]{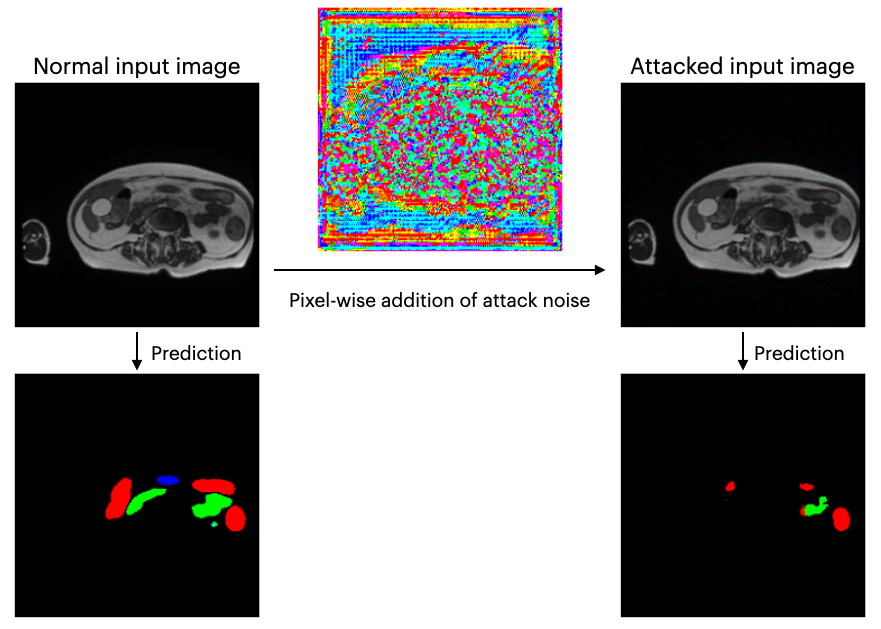}
    \caption{An example of adversarial attack on the medical MRI image segmentation model.}
    \label{fig:advattackfig}
\end{figure}

Most of the current research on adversarial attacks and related defense techniques only involve image classification  \cite{9098628} \cite{BACKDOORVIT}, which outputs a confidence score for its classification. In contrast, little research has been done adversarial attacks on image segmentation tasks. Further, most papers\cite{8237562} \cite{1048550} \cite{171109856} exploring adversarial attacks on image segmentation datasets have been done using the ImageNet \cite{IMAGENET} dataset. However, those works haven't explored medical datasets, which are proven to be more vulnerable to adversarial attacks \cite{AdvMedical}. Other works using medical imaging segmentation datasets also do not account for MRI datasets, which could reveal more detailed features of soft tissues or nerves \cite{fayad_2021}. In addition, most of such works have focused on testing lightweight models that are no longer widely used in modern applications \cite{ADVLHTMDLS} \cite{QUERYBLCBOX} \cite{kwon_2021}, even though more recent models have proven to be increasingly supceptible to adversarial attacks \cite{PROPERTIESNN}. An example of an adversarial attack on an MRI image segmentation model is shown in Figure \ref{fig:advattackfig}. 

In light of the grave danger that poisoned MRI data could pose and evident lack of research in this area, we test the susceptibility of modern MRI image segmentation models to a popular white-box adversarial attack method called Fast Gradient Sign Method (FGSM). The main motivation behind this research is to raise awareness in the academic community on the security of the MRI image segmentation models. We summarize the main contributions below:

\begin{enumerate}
    \item Through experimenting with different losses to conduct the FGSM attack, we show that using the BCE loss to conduct the attack leads to greater success than using the default loss as suggested by the FGSM paper's author.
    \item We show that having more parameters does not necessarily make the model more vulnerable to attacks. 
    \item We show that FGSM can effectively mislead modern image segmentation models models.
\end{enumerate}

\section{Approaches}

In this section, we will describe the models we selected in detail, the reasons behind those choices, the dataset we used, our adversarial attacking strategy, and our training hyperparameters.

\subsection{Model Architecture}

In this section, we will introduce U-Net and U-Net++. We will also describe why we chose to use VGG16, ResNeXt-101, and EfficientNet-B7 as our backbones.

\subsubsection{U-Net}

\begin{figure}
    \centering
    \includegraphics[scale=0.22]{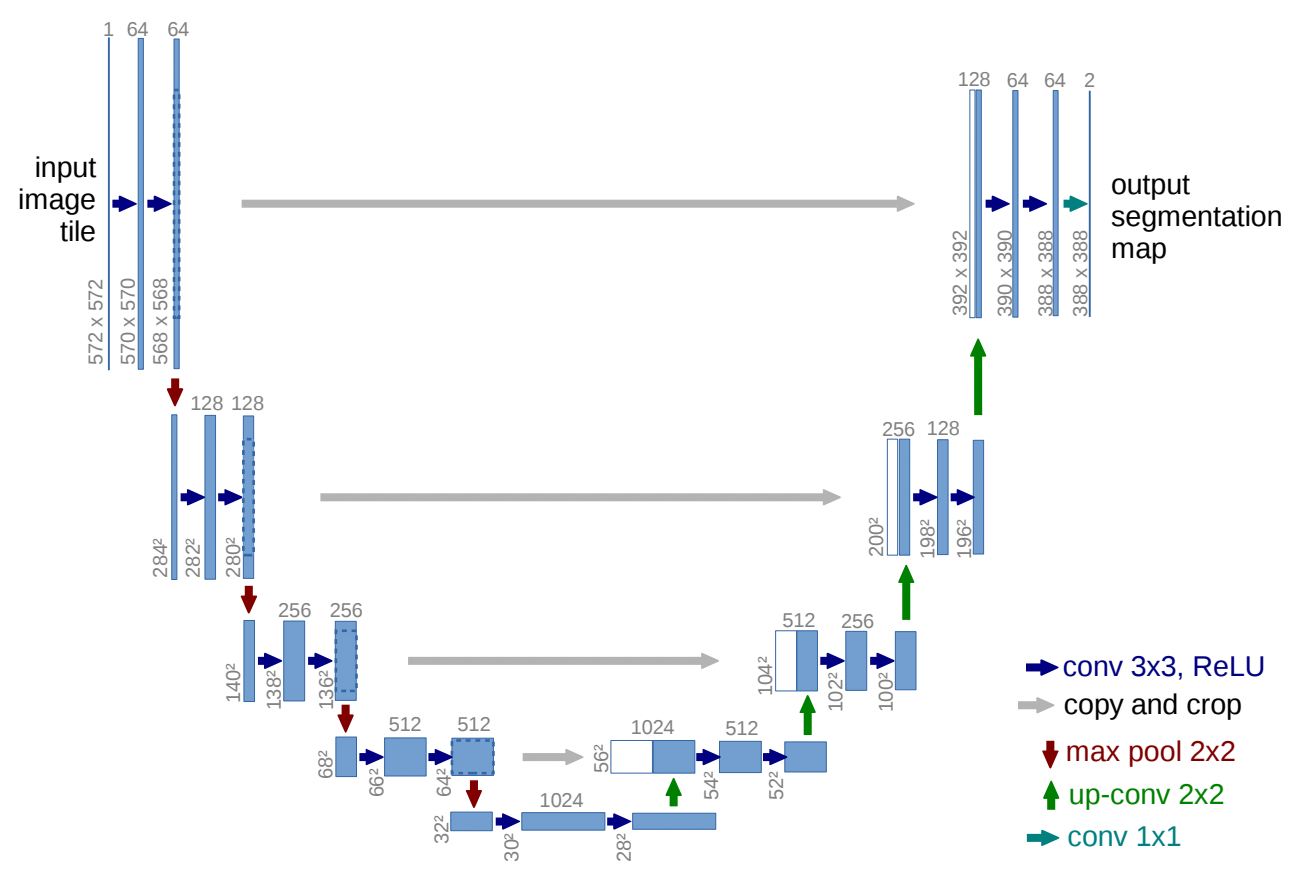}
    \caption{U-Net Architecture by Ronneberger et al. Here the architecture assumes the input is 572 $\times$ 572}
    \label{fig:unet}
\end{figure}

U-Net, introduced by Ronneberger et al., is one of the most commonly used image segmentation architectures for biomedical imaging \cite{UNet}. U-Net is a u-shaped network containing a down-sampling path, a bottleneck, and an up-sampling path (Figure \ref{fig:unet}). During each level of the down-sampling path, the dimension of the image is contracted by the max pooling layer, yet the number of feature channels is expanded by a factor of two, which allows the network to learn global features better. During each level of the up-sampling path, the output of the previous layer is concatenated with the output from the same level’s down-sampling path to simultaneously fuse the global and local features necessary for segmentation.

Among four of our models, three of them are U-Net based. 

\begin{itemize}
    \item {\textbf{U-Net}: The first model is the basic U-Net with no modified backbones. We use this model as our baseline.}
    \item {\textbf{U-Net w/ ResNeXt-101}: The second model is based on the U-Net's architecture, but the down-sampling path is replaced by layers from a pre-trained ResNeXt101 32x8d model \cite{RESNEXT}. ResNeXt101 is a 101-layer variant of ResNeXt, which achieved second place in the ILSVRC 2016 classification competition \cite{DBLP:journals/corr/XieGDTH16}. }
    \item {\textbf{U-Net w/ EfficientNet-B7}: The third model we use is also based on the U-Net's architecture, but the encoder layers are replaced by layers from a pre-trained EfficientNet-B7 model \cite{efficientnet}. EfficientNet-B7 is a complex variant of the EfficientNet family, which achieves SOTA efficiency by outperforming most models within its domain with much higher efficiency \cite{efficientnet}. Today, EfficientNet-B7 is extensively used in industry and in medical imaging competitions \cite{RSNA_2nd_effnet} \cite{tract_tumor1st} \cite{tract_tumor2nd} \cite{tract_tumor3rd}}.
\end{itemize}


\begin{figure}
    \centering
    \includegraphics[scale=0.22]{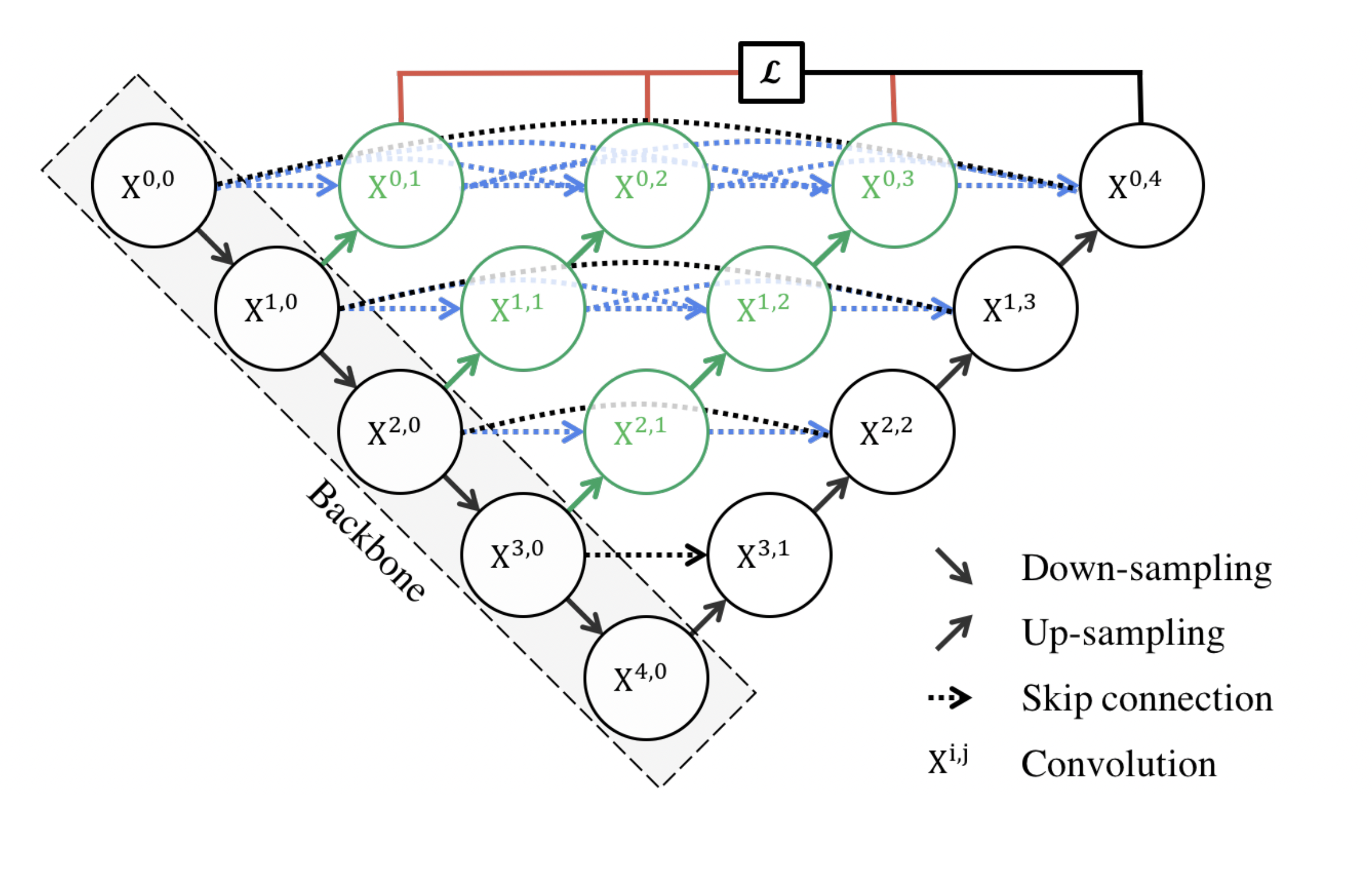}
    \caption{UNet++ Architecture by Zhou et al. UNet++ introduced dense skip connections as highlighted in \textcolor{blue}{blue arrows} and \textcolor{green}{green arrows} shown in the graph. The model also introduced deep supervision as indicated by $L$, but in our paper, this is excluded during training.}
    \label{fig:unet2p}
\end{figure}

\subsubsection{UNet++}

UNet++ by Zhou et al. introduces an important innovation to the U-Net network — dense skip connections \cite{UNET2P}. As shown in Figure \ref{fig:unet2p}, instead of naively concatenating the feature map from each level of the encoder layer $X^{1, 0}, X^{2, 0}, X^{3, 0}$ to the feature maps of the corresponding decoder layer $X^{2, 2}, X^{1, 3}, X^{0, 4}$, intermediate dense skip connections are introduced, as shown in green colored nodes in Figure \ref{fig:unet2p}. Dense skip connections allow the model to learn faster because the image representations are richer, and the semantic gap is smaller.

The last trained model is U-Net++ based.
\begin{itemize}
    \item {\textbf{UNet++ w/ EfficientNet-B7}: The fourth model we used is based on the U-Net++'s architecture and has an EfficientNet-B7 backbone \cite{efficientnet}. We choose EfficientNet-B7 because our prior experiments show that a pre-trained EfficientNet-B7 has a superior performance when used as an encoder.}
\end{itemize}

\subsection{Datasets}
All the models we use in our experiements were trained using University of Wisconsin-Madison's gastro-intestinal tract (UW-Madison GI Tract) MRI image segmentation dataset \cite{kaggle_2021}, which is publicly available on Kaggle. The dataset is made up of 272 workable 3D scans and 38208 images that are black-and-white. Segmentation masks are encoded in the run-length encoding (RLE) format. 

There are three classes in this dataset: large bowel (14,017 images), small bowel (11,129 images), and stomach (8,558 images). Instances chosen for training and testing datasets were carefully picked to ensure they all have a similar distribution. To prevent the data leakage problem, slices from individual scans were grouped together and together either in the training set or the testing set. The detailed dataset distribution is shown in Table \ref{tab:dataset}. 


\subsubsection{Pre-processing}
During pre-processing, all pixels were normalized to range from $[0, 1]$, and all images are resized to $224 \times 224$. ($224 = 32\times7$).


\begin{table}[]
\caption{Detailed dataset distribution.}
\begin{tabular}{cccc}
                  & Training (Slices) & Testing (Slices) & Total (slices) \\
Large Bowel Tumor & 12,698                     & 1,319                     & 14,017          \\
Small Bowel Tumor & 9,955                      & 1,174                     & 11,129          \\
Stomach Tumor     & 7,611                      & 947                      & 8558           \\
Total             & 34,432 (90\%)              & 3,776 (10\%)              & 38,208         
\end{tabular}
\label{tab:dataset}
\end{table}

\subsection{Model Training}
In this section, we will share the training parameters that we used to conduct our experiments. All of our models were trained without sufficient fine-tuning because we prioritized analyzing the impact of adversarial attack over gaining the best performances on normal input images for all models.

\subsubsection{Hardware and Software}
All four of our models were trained on one Nvidia A6000 (48GB) instance with 14vCPUs and 100 GiB RAM.

\subsubsection{Hyperparameters}
All models were trained using AdamW with an initial learning rate of $3e-4$ and a weight decay of $1e-3$. We also used a cosine annealing learning rate scheduler with max iterations of 7,081 and a maximum learning rate of $3e-4$. We trained all our models for 15 epochs with early stopping. We used a batch size of 64 and applied data augmentation on training images in the form of shifting, scaling, and deformation.

For our U-Net models, instead of having the number of kernels starting at 64 and growing by a factor of two until 1024, we modified it to have a kernel number beginning at 16 and ending at 256 because to make the model more efficient during training.

\subsubsection{Loss Function}
All of the models in this paper were trained using a \textbf{hybrid loss} that combined the Dice loss and Focal loss, which helped to deal with class imbalance and improve the performance of our models \cite{dicefocal}. 

Let's define $y \in \{0, 1\}$ as the ground truth mask and $\hat{y} \in [0, 1]$ as the predicted mask. 

\textbf{Dice loss (DL)} \cite{SEGLOSSSURVEY} is defined in Formula \ref{DICELOSS}. The Dice loss is basically $1 - DSC$. DSC is expanded in section \ref{EVALMET} We add $10^{-6}$ in the numerator and denominator to avoid division by zero.

\begin{equation}
\label{DICELOSS}
DL =  1 - 2 \times \frac{y \hat{y} + 10^{-6}}{y+\hat{y}+10^{-6}} 
\end{equation}

\textbf{Focal loss (FL)} \cite{F0CALLOSS} is defined in Formula \ref{FLFORMULA}. Focal loss improves \textbf{binary cross entropy (BCE) loss} \cite{CELOSS} by dealing with the imbalanced dataset problem. We derive the focal loss formula firstly by deriving the binary cross entropy loss formula, as shown in Formula \ref{BCELOSS}. 

We define $C$, $W$, and $H$ to indicate the number of channels, the height, and width of the image. $c \in [0.. C)$, $i\in [0.. W)$, and $j \in [0.. H)$ are indexes. For example, $y_{c,i,j}$ means the pixel value of the mask $y$ at channel index $c$, width index $i$, and height index $j$.

So, we define $\hat{y}^t$ that for each pixel of $y$,

\begin{equation}
    \hat{y}^t_{c,i,j}=
    \begin{cases}
    \hat{y}_{c,i,j}, & \text{if }y_{c,i,j}=1\\
    1-\hat{y}_{c,i,j}, & \text{if }y_{c,i,j}=0
    \end{cases}
\end{equation}

Therefore, binary cross entropy loss for image segmentation can be defined in the formula below.

\begin{equation}
\label{BCELOSS}
    BCE(y,\hat{y}) = - \sum_{c=0}^{C} \sum_{i=0}^{W} \sum_{j=0}^{H} \log(\hat{y}^t_{c,i,j})
\end{equation}

Focal loss adds a modulating factor $(1-\hat{y}^t)^{\gamma}$ to BCE. $\gamma$ is a tunable hyperparameter. Setting $\gamma > 0$ will differentiate focal loss from binary cross entropy loss. Setting $\gamma > 1$ would make the model less sensitive to class imbalance, and setting $1 > \gamma > 0$ would make the model more sensitive to class imbalance.

\begin{equation}
\label{FLFORMULA}
    FL(y, \hat{y})= - \sum_{c=0}^{C} \sum_{i=0}^{W} \sum_{j=0}^{H} (1-\hat{y}^t_{c,i,j})^{\gamma} \cdot \log(\hat{y}^t_{c,i,j})) 
\end{equation}


For all our experiments we set $\gamma=2$.


\subsection{Fast Gradient Sign Method}

Goodfellow et al. proposed the Fast Gradient Sign Method (FGSM), which would generate adversarial inputs by nudging the input in the direction of the gradient with respect to the input space. \cite{FSGM}.

\begin{figure}
    \centering
    \includegraphics[scale=0.148]{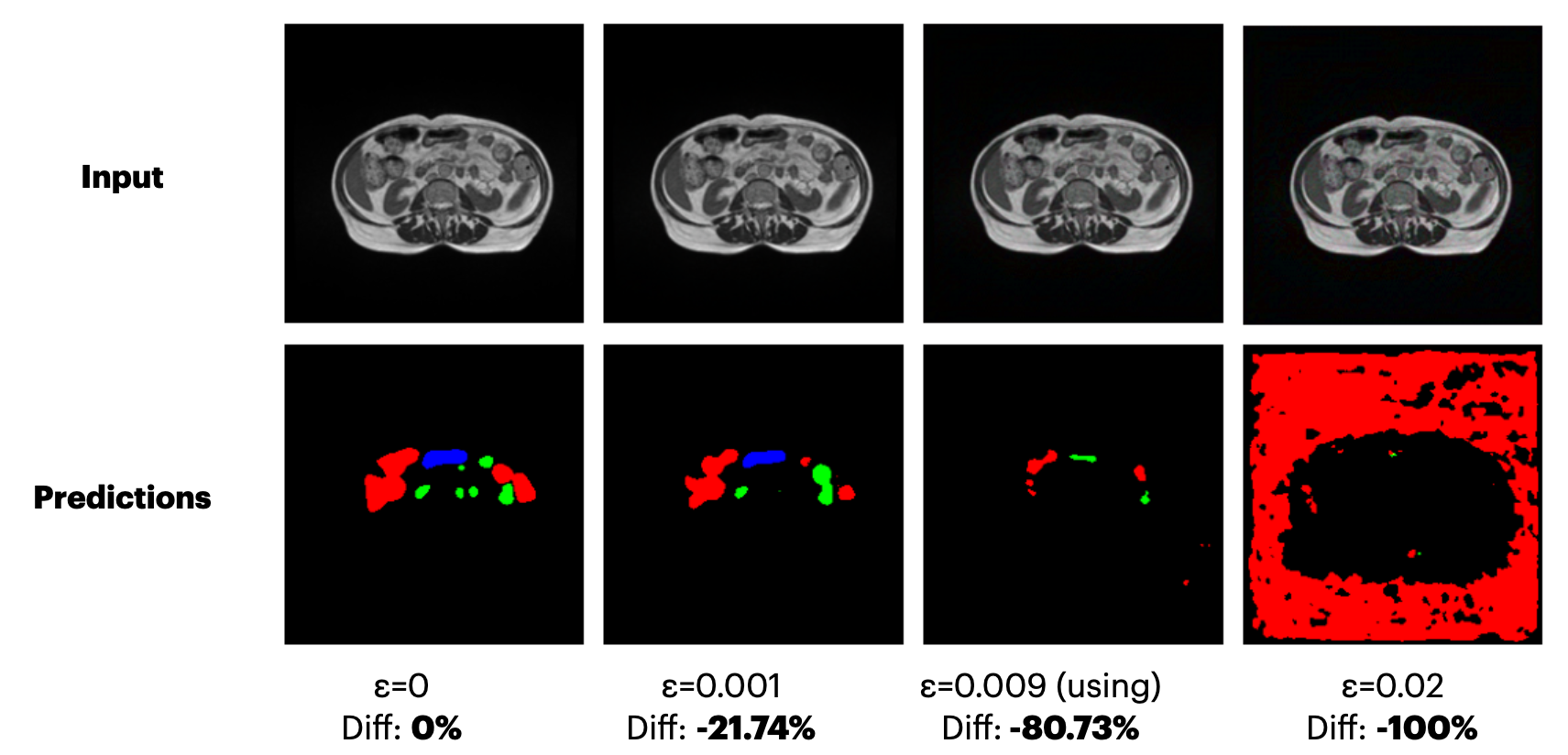}
    \caption{Input images and predictions with various $\epsilon$ values for our U-Net++ Model. Diff value is measured by the clean input metric score minus the poisoned input metric score divided by the clean input metric score again. Therefore, the more drastic the difference is, the more successful the attack would be.}
    \label{fig:diff_eps}
\end{figure}

The FGSM attacking formula is provided below, where $adv_y$ is the adversarial image, $\theta$ is the parameters of the model, $x$ is the input to the model, $y$ is the prediction of the model, $J(\theta, x, y)$ is the loss function of the model, $sign$ is the sign of the gradient with respect to the pixels used in the back-propagation stage, and $\epsilon$ is the multiplier of the noise that can be tuned to achieve a balance between stealthiness and effectiveness. Figure \ref{fig:diff_eps} shows under different $\epsilon$ values, what do the input image and the predicted mask look like. It shows that the higher the $\epsilon$ value, the more successful the attack will be, but the input image will loss the stealthiness as the attack noises would become gradually visible.

\begin{equation}
Adv_y = x + \epsilon * sign( \nabla_x J(\theta, x, y) )
\end{equation}
FGSM is a simple yet robust adversarial attack. The attack is also illustrated in Figure \ref{fig:advattackfig}. In this paper, we compared the performance of all four of our segmentation models before and after the FGSM attack. In all experiments, we use $\epsilon=0.009$ because, as shown in Figure \ref{fig:diff_eps}, $\epsilon=0.009$ achieves both stealthiness and effectiveness.

In the original FGSM paper, authors Goodfellow, Shlens, and Szegedy, suggested that $J(\theta, x, y)$ should be the loss function used to train the network \cite{FSGM}. However, during our experiments, we found that using the original loss function led to less effective attacks than using an alternative loss. For our experiments, we found that using binary cross entropy (BCE) loss function led to significant improvement in attacking success.

\section{Results}

The results are shown in Table \ref{tab:result}. Since there are some false negative ground truth masks in this dataset, we only test our models' performance using the MRI slices that have segmentation masks. 

For our experiments, U-Net w/ ResNeXt-101 has the most number of parameters, followed by U-Net++ w/ EfficientNet-B7 and U-Net w/ EfficientNet-B7 models.

\begin{figure}
    \centering
    \includegraphics[scale=0.153]{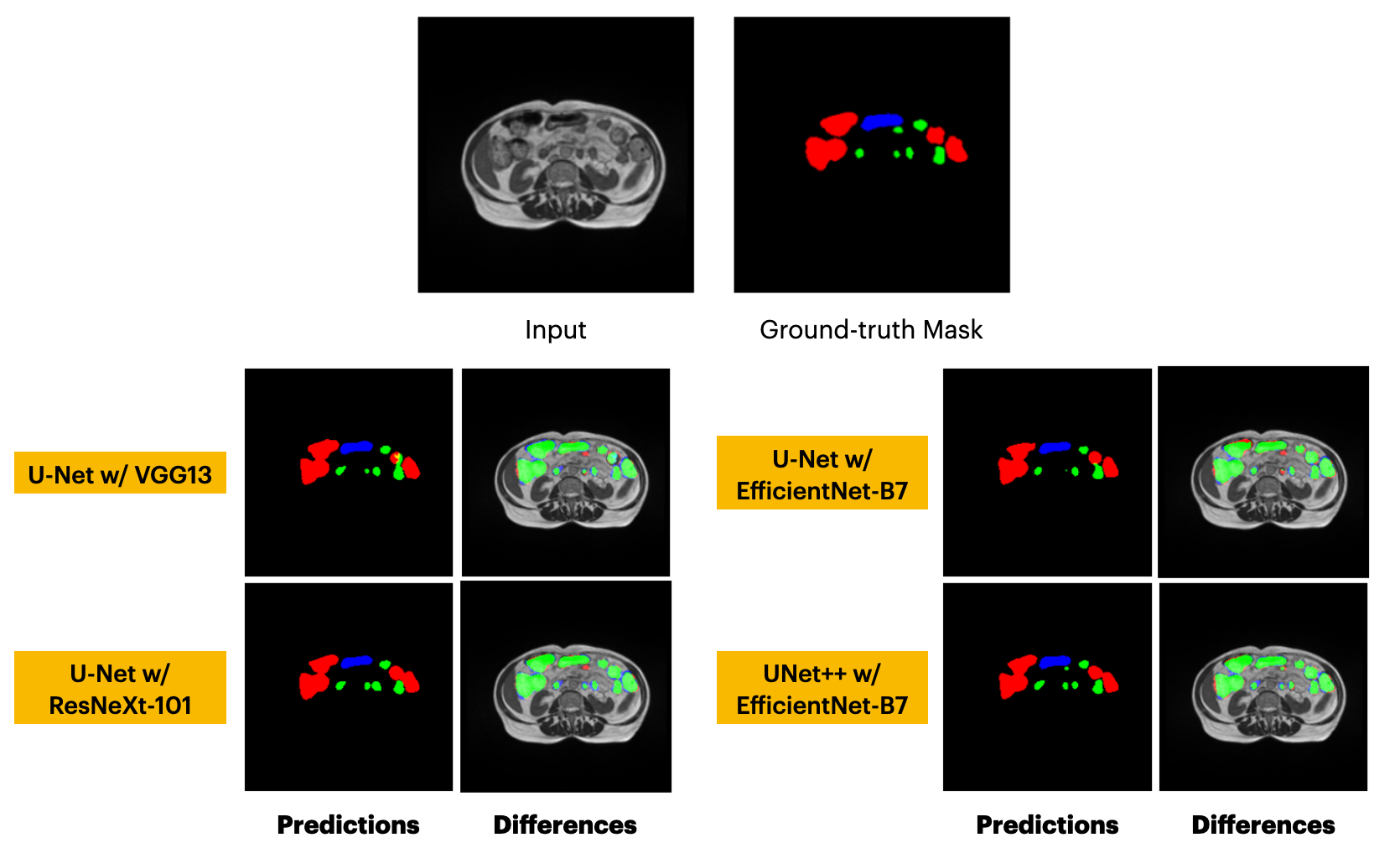}
    \caption{Comparison of the predicted masks of four models. Three colors in the predictions columns indicate three different objective classes. Model differences to the ground-truth mask are also illustrated. For the differences, true positive is highlighted in \textcolor{green}{green}, false positive is highlighted in \textcolor{blue}{blue}, and false negative is highlighted in \textcolor{red}{red}.}
    \label{fig:model_preds}
\end{figure}

For normal inputs, U-Net++ with EfficientNet-B7 and U-Net with EfficientNet-B7 models are the most successful model among all with U-Net with EfficientNet-B7 model performing slightly worse. Their predictions based on a normal image input and their differences to the original mask are illustrated in Figure \ref{fig:model_preds}. All four models predicted the masks well with few false negatives and false positives.

\begin{figure}
    \centering
    \includegraphics[scale=0.215]{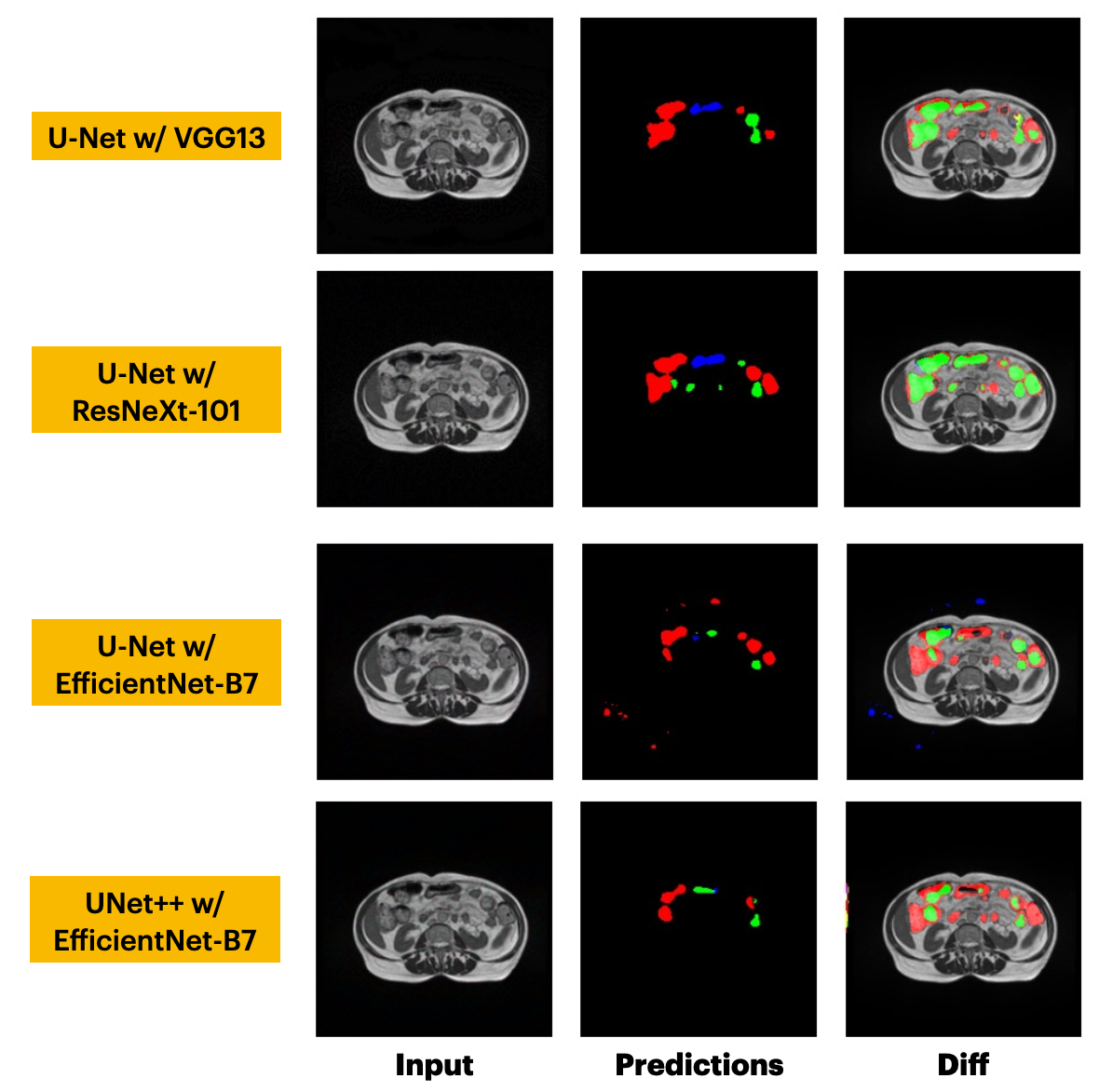}
    \caption{Adversarial attacks on four models. The first column contains the poisoned inputs. The second column contains predictions using the poisoned inputs. The third column indicates their differences. However, it should be noted that for the first two models, we used a higher epsilon value, $\epsilon=0.015$, for illustration purposes because the first two models are very resilient to the attack. we also talk it briefly in future work section \ref{futurework}. For the differences, true positive is highlighted in \textcolor{green}{green}, false positive is highlighted in \textcolor{blue}{blue}, and false negative is highlighted in \textcolor{red}{red}.}
    \label{fig:4x3adv}
\end{figure}

However, all four models were all significantly impacted by FGSM, instantly making them unreliable for doctors, as shown in Figure \ref{fig:4x3adv}.

\begin{figure}
    \centering
    \includegraphics[scale=0.15]{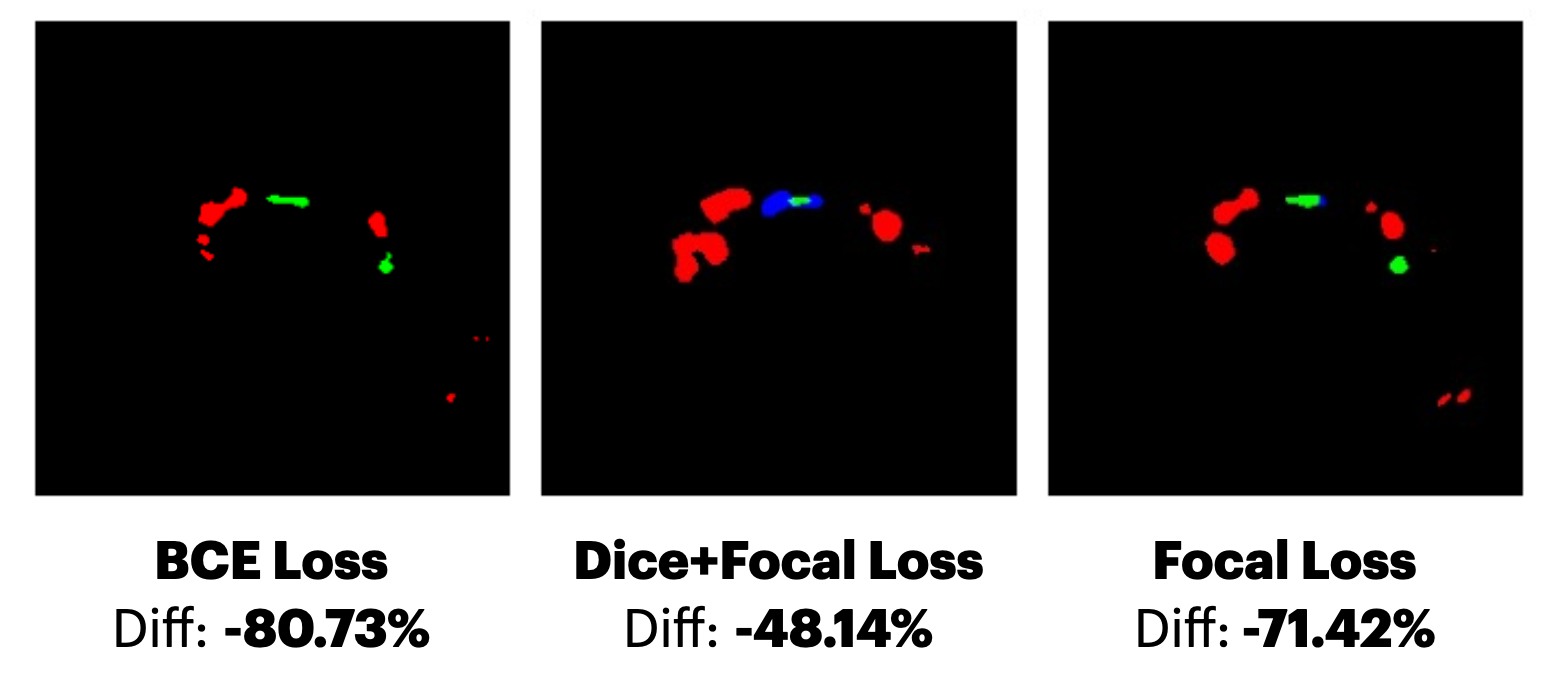}
    \caption{Prediction masks showing the differences between the use of binary cross entropy (BCE) loss, hybrid loss combining dice and focal loss, and focal loss for an FGSM attack. Attacks were done on the U-Net++ w/ EfficientNet-B7 model for the same image. Diff value is measured by the clean input metric score minus the poisoned input metric score divided by the clean input metric score again.}
    \label{fig:my_label}
\end{figure}

Since FGSM attacks require a loss function to derive the signs of the gradients, we also tested out three different loss functions to see how the attack success varies. It turned out that BCE had the highest success rate, despite the fact that our original model was trained on a hybrid loss of focal and dice loss. The comparison between three loss functions is shown in Figure \ref{fig:my_label} FGSM achieves the highest success in U-Net++ w/ EfficientNet-B7 model in all three loss functions.

Our results also imply a negligible correlation between the number of parameters and attack success rate. However, it is worth noting that U-Net w/ EfficientNet-B7 and U-Net++ w/ EfficientNet-B7 models, which had the best performances for clean inputs, were the most vulnerable to the FGSM attack. U-Net++ w/ EfficientNet-B7 model has not only the best performance but also the highest attacking success rate.

In addition, even though the authors of the FGSM paper suggested to use the cost function used to train the model to conduct the attack, empirical evidence suggests that it is not true in this case. The original cost function used to train all four models is the hybrid loss combing focal loss and dice loss. As shown in Table \ref{tab:result}, combining focal loss and dice loss to conduct the attack received the lowest attack success, yet using binary cross entropy (BCE) loss to conduct the attack yielded significantly higher attacking success.


\subsection{Evaluation Metric}
\label{EVALMET}

\noindent
\textbf{Dice Similarity Coefficient (DSC):} To evaluate the models' performance on testing data before and after applying adversarial noise, we used DSC that would measure the effectiveness of the overlap between the ground truth and predicted mask. DSC is bounded between $-1$ and $+1$. \cite{DICECOEF}. Its formula is defined below. $y$ and $\hat{y}$ are the ground-truth mask and the predicted mask.

\begin{equation}
DSC = 2 \times \frac{|y \cap \hat{y}|}{|y| + |\hat{y}|}    
\end{equation}

\noindent
\textbf{Attacking Success (AS):} To evaluate the effectiveness of the adversarial attack, we created a metric that would measure the percentage change in DSC, as shown in Formula \ref{ASMETRIC}. The resulting value is a percentage between 0\% and 100\%, and higher the AS the more successful the attack is.

\begin{equation}
\label{ASMETRIC}
    AS = \frac{\text{DSC Before Attack} - \text{DSC After Attack}}{\text{DSC Before Attack}}
\end{equation}


\begin{table}[]
\centering
\caption{Comparison of the performance before and after a FGSM attack was done on our models: U-Net with VGG13  (VGG U-Net), U-Net with ResNeXt-101 (ResNeXt U-Net), U-Net with EfficientNet-B7 (EffB7 U-Net), and U-Net++ with EfficientNet-B7 (EffB7 U-Net++) on the GI Tract dataset. All measured in Dice Similarity Coefficient (DSC) score. Attacking successes of three losses used for FGSM attack were compared. The most successful attacks are highlighted in \textbf{bold}.}
~\clap{
\begin{tabular}{cccccc}
\hline

\multicolumn{1}{l}{}       & \multicolumn{1}{l}{}                    & \multicolumn{1}{l|}{} & \multicolumn{3}{c|}{\textbf{FGSM Using Loss}}             \\ \hline
\textbf{Model Names}       & \multicolumn{1}{l}{\textbf{Parameters}} & \textbf{Normal}       & \textbf{BCE}     & \textbf{Focal+Dice} & \textbf{Focal}   \\ \hline
VGG U-Net            & 18.44M                                  & 0.7509               & 0.4063          & 0.5772             & 0.5105          \\
ResNeXt U-Net       & 95.76M                                  & 0.7841               & 0.3873          & 0.6197             & 0.5560          \\
EffB7 U-Net   & 67.10M                                  & 0.7994               & 0.4576          & 0.5348             & 0.5097          \\
EffB7 U-Net++ & 68.16M                                  & \textbf{0.8024}      & \textbf{0.3750} & \textbf{0.4705}    & \textbf{0.4330} \\ \hline
\end{tabular}
}

\label{tab:result}
\end{table}

\section{Discussions and Conclusions}

In this paper, we trained four advanced image segmentation models from the U-Net family and examined the efficacy of FGSM for poisoning MRI data to understand how vulnerable they are to adversarial attacks. We observe that all the models in this paper are heavily impacted by FGSM, stressing an urgent need to enact serious security measures under professional environments. In addition, we observe that even though the FGSM paper suggests using the loss function used to train the model, using binary entropy loss as an alternative to generate attacking noises under this context has consistently demonstrated better attacking success rates. Lastly, we observe that having more parameters does not necessarily imply the vulnerability of the model to adversarial attacks.

\subsection{Future Work}
\label{futurework}

In future studies, we will test the adversarial robustness of more of U-Net models using various attacking methods. Doing this would give the academic community a more complete sense of what models are more susceptible to adversarial attacks, and what types of adversarial attacks are likely be successful. 

During our experiment, we also noted that some models are resilient to certain types of images, while others are not. Specifically, we found out image luminosity seems to be an important factor. We will conduct more experiments to find out if there is a relationship between image brightness and attack successes for certain models.

Finally, we will also test out the effectiveness of adversarial attacks by ensembling all our trained models.

\section*{Acknowledgment}
We thank our friends, parents, and teachers for constantly supporting us throughout this process.

\bibliographystyle{ieeetr}
\bibliography{bib}


\end{document}